%% file: mad09_pioppi.tex
\documentclass[12pt]{article}
\input{symbols}
\usepackage{empheq} 
\usepackage{epsfig}
\usepackage{xspace}
\usepackage{subfigure}
\input econfmacros.tex
\def\Title#1{\begin{center} {\Large {\bf #1} } \end{center}}

\begin{document}

\Title{Search for SUSY at LHC in the first year of data-taking.}
  \begin{abstract}
If Supersymmetry would manifest itself at a low mass scale it might be found 
already in the early phase of the LHC operation. Generic signatures for 
Supersymmetry in pp-collisions consist of high jet multiplicity, 
large missing transverse energy as well as leptons in the final state. 
The CMS search strategy and prospects for a SUSY discovery 
in the first year of data-taking is reviewed.
  \end{abstract} 
\bigskip\bigskip


\begin{raggedright}  

{\it Michele Pioppi\index{Pioppi, M.}\footnote{also with INFN Perugia, Italy} on behalf of the CMS collaboration\\
Imperial College London\\
South Kensington Campus, \\
London SW7 2AZ, United Kingdom} \\

\bigskip\bigskip
\end{raggedright}

\input{intro}

\input{jet}
\input{photon}
\input{lepton}
\input{conc}

 \end{document}

%% file: symbols.tex
\newcommand{\ra}{\ensuremath{\rightarrow}}
\newcommand{\znunu}{\ensuremath{{\text Z} \ra \nu\bar{\nu}}}

\newcommand{\pt}{\ensuremath{p_{ \mathrm{T}} }}
\newcommand{\et}{\ensuremath{E_{ \mathrm{T}} }}
\newcommand{\ptjf}{\ensuremath{p_{\rm T}^{ {\rm j}_1} }}
\newcommand{\ptjs}{\ensuremath{p_{\rm T}^{ {\rm j}_2} }}
\newcommand{\ptjt}{\ensuremath{p_{\rm T}^{ {\rm j}_3} }}

\newcommand{\alt}{\ensuremath{\alpha_{ \text{T}}}}

\newcommand{\ttNew}{\ensuremath{\rm{t}\bar{\rm{t}}}}

\newcommand{\pbinv} {\mbox{\ensuremath{\,\text{pb}^\text{$-$1}}}\xspace}
\newcommand{\MET}{\ensuremath{E_{\mathrm{T}}^{\mathrm{miss}}}\xspace}
\newcommand{\GeV}{\ensuremath{\,\text{Ge\hspace{-.08em}V}}\xspace}
\newcommand{\TeV}{\ensuremath{\,\text{Te\hspace{-.08em}V}}\xspace}
\newcommand{\GeVc}{\ensuremath{{\,\text{Ge\hspace{-.08em}V\hspace{-0.16em}/\hspace{-0.08em}c}}}}

%% file: econfmacros.tex
\def\beq{\begin{equation}}
\def\eeq#1{\label{#1}\end{equation}}
\def\eeqn{\end{equation}}

\def\beqa{\begin{eqnarray}}
\def\eeqa#1{\label{#1}\end{eqnarray}}
\def\eeqan{\end{eqnarray}}

\let\bar=\overbar

\def\Dslash{\not{\hbox{\kern-4pt $D$}}}
\def\dslash{\not{\hbox{\kern-2pt $\del$}}}

\def\msb{{\bar{\ssstyle M \kern -1pt S}}}



%% file: intro.tex
\section{Introduction}
\label{sec:intro}
The Large Hadron Collider (LHC)~\cite{LHC} at CERN opens a new energy regime
offering a very exciting discovery potential for physics beyond the Standard Model. 
In this paper the search for Supersymmetry (SUSY) in the startup scenario at $\sqrt{s}=$ 10 \TeV\ is 
discussed.
Supersymmetry exists in many theories beyond the Standard Model. Two of the best studied 
braking mechanisms are the Minimal SuperGravity (mSUGRA) and Gauge Mediated Symmetry Breaking (GMSB).
Benchmark points
have been defined in the framework  of mSUGRA and GMSB to study various experimental SUSY signatures. 
In the CMS experiment~\cite{PTDRI} SUSY analyses are organized according to topologies, e.g. number
of leptons, photons and jets in the final state, which arise from SUSY cascades.
A detailed definition of the benchmark points can be found here~\cite{PTDRII}.

The paper is organized as follows:
in Section~\ref{sec:jet} the SUSY search in multijet final state is discussed~\cite{SUS0901}. 
Searches for new physics in diphoton~\cite{SUS0904} and dilepton~\cite{SUS0902} final state are 
described in section~\ref{sec:phot} and~\ref{sec:lept} respectively.

%% file: jet.tex
\section{Search for SUSY in fully hadronic final state}
\label{sec:jet}
This section describes a search strategy for a possible discovery of
SUSY signatures with the CMS detector 
at the LHC using exclusive $n$-jet events ($n = 2 \dots 6$).
The event topology under investigation consists of $n$ high-\pt\ jets
and two invisible neutralinos which lead to a missing energy
signature. The high-\pt\ jets are produced in the decay chains of the initially
produced heavy squarks and gluinos.
The main aim of the analysis is to develop a
robust measurement technique suitable for the early physics
data at the LHC and stable with respect to jet energy mismeasurements. 
Before applying any event selection, multijet production from 
QCD is the dominant process, where missing energy is introduced through 
jet mismeasurements.

\subsection{Trigger and selection}
The benchmark points of mSUGRA LM0 and LM1 are used to estimate the trigger efficiency 
for signal events. Both signal points have 100$\%$ efficiency after all cuts 
for the single jet trigger HLT\_Jet110 
(one jet with corrected jet transverse momentum $>$ 110~\GeVc).

Hadronic jets are reconstructed from calorimeter energy deposits which are
clustered using an iterative cone algorithm with 
$R=0.5$~\cite{JetPerformancePAS}.
Furthermore, these jets are required to have a transverse momentum  greater than 50~\GeVc,  
pseudo-rapidity $|\eta| < 3.0$, and 
an electromagnetic fraction $F_{\rm em} < 0.9$. 
The transverse momentum of the leading jet
and second leading jet need to exceed 100~\GeVc\ and 
the pseudo-rapidity of the leading jet  is required to be smaller than two.
Based on the jets defined above 
two additional variables are defined: $H_{\rm T}$
as the scalar sum over the transverse momenta of the selected jets in an event, 
$H_{\rm T}$ = $\sum_i \pt^{j_i}$ and the missing transverse momentum of the event 
calculated as $\vec{H}_{\rm T}^{\rm miss} = -\sum_i \vec{\pt}^{j_i}$.  
In order to reduce background events from SM processes 
$H_{\rm T}$ is required to be greater than 350~\GeVc. \\
All events are rejected where either an isolated  muon~\cite{muonReco} with \pt\ $>$ 10~\GeVc\  
or an isolated electron~\cite{PASelectron} with  \pt\ $>$ 10~\GeVc\ or
photons~\cite{PTDRII} with \pt\ $>$ 25~\GeVc\
or jets with  \pt\ $>$ 50~\GeVc\ that does not fulfill the other criteria 
($|\eta| < 3$ or $F_{\rm em} < 0.9$)
are found.

\subsection{Analysis method and results}
In the following a kinematic variable (\alt) is used that allows separation of signal events with 
real missing energy from QCD events in which missing energy is created by jet energy mismeasurements.\\
In the dijet case ($n = 2$) transverse momentum conservation requires
the \pt\ of the two jets in QCD events to be of equal magnitude and
back-to-back in the azimuthal angle $\phi$.  The variable \alt, first
introduced in Ref.~\cite{cms-pas-sus-08005}, exploits exactly this requirement. 
It is defined as
\begin{equation}
 \label{eq:altsimple}
\alt = \et^{j2}/\text{M}_\text{T}  \quad
\end{equation}
where $\et^{j2}$ is the transverse energy of the second leading jet in the event and $\text{M}_\text{T}$ 
is defined as
\begin{equation}
  \label{eq:alphatmassless}
  \text{M}_\text{T} = \sqrt{ \left( \sum_{i=1}^n \et^{j_i} \right)^2 - \left( \sum_{i=1}^n p_x^{j_i} \right)^2 - \left( \sum_{i=1}^n p_y^{j_i} \right)^2} = \sqrt{H_{\rm T}^2 - (H_{\rm T}^{\rm{miss}})^2} \quad ,
\end{equation}
and $n=2$ in the dijet case.
For a well measured QCD dijet event, $\et^{j2} = 0.5 \times H_T$ and $H_{\rm T}^{\rm miss} = 0$, 
thus \alt\ is exactly 0.5.

To define \alt\ for more than two jets the $n$-jet system is reduced down 
to a two-jet system by combining jets into two pseudo-jets. The \et\ of the pseudo-jets
is calculated as the scalar sum of the contributing jet \et. All possibilities of how $n$ 
jets can be combined into two are tested and the combination is chosen where the resulting 
pseudo-jet \et\ are most similar, \textit{i.e.}, for which the difference 
$\Delta H_{\rm T} = \et^{pj_1} - \et^{pj_2}$ is minimal.
For $n$ jets, \alt\ is then obtained in the same way as in Eq.~\ref{eq:altsimple}.

In the event selection $H_{\rm T}$ is required
to be greater than 350~\GeVc\ which is well above the transverse
momentum threshold of 50~\GeVc\ for a single jet. However several jets
below that threshold could still lead to a considerable amount of
ignored momentum in the event. For that reason the $H_{\rm T}^{\rm miss}$ 
determined using all jets having a \pt\ larger than 30~\GeVc\ ,  $H_{\rm T}^{\rm miss}(\mbox{jet}\, p_{\rm T} > 30~\GeVc)$, 
is calculated and compared to the $H_{\rm T}^{\rm miss}$ determined from the selected
jets only, $H_{\rm T}^{\rm miss}(\mbox{selected jets})$. The ratio 
\begin{equation}
R(H_{\rm T}^{\rm miss}) = H_{\rm T}^{\rm miss}(\mbox{selected jets})/H_{\rm T}^{\rm miss}(\mbox{jet}\, p_{\rm T} > 30~\GeVc)
\end{equation}
can be used to single out events where
the inclusion of lower momentum jets does significantly improve the
balance of the event. 
If the missing transverse energy 
($H_{\rm T}^{\rm miss}$) is increased by 25\% due to the fact that the transverse 
momentum threshold of the selected jets is 50~\GeVc\ and not 30~\GeVc\, the event is 
rejected, thus $R(H_{\rm T}^{\rm miss})$ is required to be smaller than $1.25$.

\begin{figure}[t]
   \begin{center}
\subfigure[Distribution of  $\alpha_T$ for dijet events.]{\label{fig:dijetA_T}\includegraphics[angle=90,width=7cm]{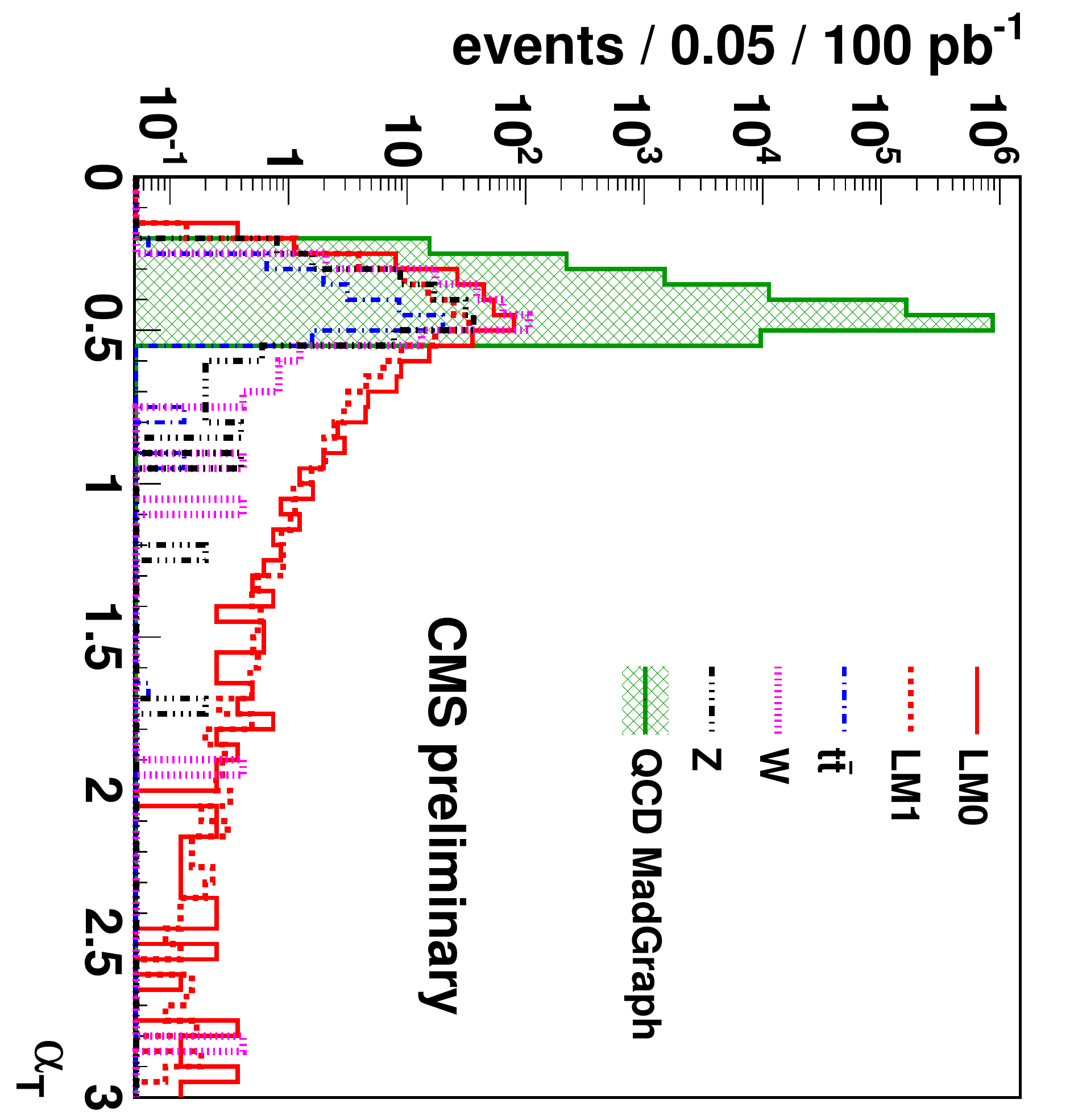}}
\subfigure[Distribution of $\alpha_T$ for events with $n=3\dots6$ jets.]{\label{fig:njetA_T}\includegraphics[angle=90,width=7cm]{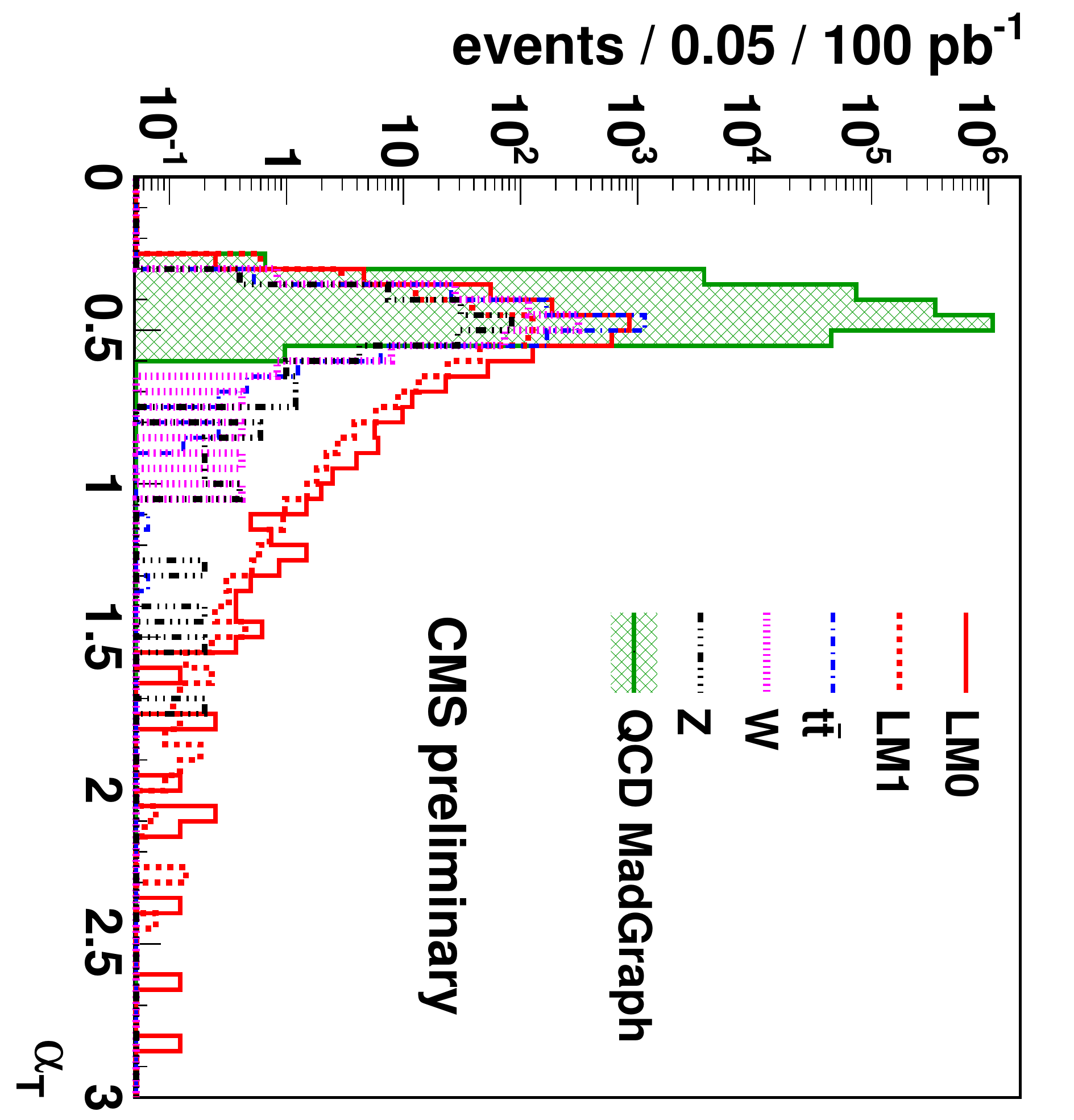}}
\end{center}
 \caption{\label{fig:alphat} $\alpha_T$ distribution.}
\end{figure}

\begin{table}[t] 
\begin{center}
\begin{tabular}{|l||c|c|c|c|c||c|c|}
\hline 

       & QCD$_{\rm MadGraph}$ &    \znunu & W$\rightarrow\nu \ell$ & \ttNew & Z$\rightarrow\ell\ell$ & LM1 & LM0 \\ \hline \hline
dijet  &0.0${+1.0}$ & 2.8$\pm$0.7 & 5.0$\pm$1.4 &  0.3$\pm$0.1  & 0.0${+0.3}$ & 52$\pm$1 & 68$\pm$3\\\hline
$n_j>$2
  & 0.9$^{+1.0}_{-0.9}$ & 10.0$\pm$1.4 & 10.4$\pm$1.7 & 8.8$\pm$0.8 & 0.3$^{+0.4}_{-0.3}$ & 116$\pm$1 & 253$\pm$6\\\hline
\end{tabular}
\caption{ \label{tab:yields}Numbers of events expected for the dijet case and for $n = 3 \dots 6$ jets case  
  for background samples (QCD, \znunu +jets, W+jets, \ttNew\, and Z+jets) and 
  the LM0 and LM1 signal points. The final numbers of selected events
  are shown after the cuts on \alt\, and $R(H_{\rm T}^{\rm miss})$.} 
\end{center}
\end{table}

The \alt\ distributions for the dijet case and the sum of $n = 3 \dots 6$ jets case 
are shown in Fig.~\ref{fig:alphat} where the requirement on $R(H_{\rm T}^{\rm miss})$ 
has already been applied. In both figures the QCD background peaks, as expected, sharply 
at a value of 0.5. To account for finite jet energy and $\phi$ resolutions events are only 
selected if $\alt$ is larger than $0.55$.  

The resulting event yields for signal and background
are summarized in Tab.~\ref{tab:yields}. 
All expected event yields correspond to an integrated luminosity of 100\pbinv.
It can be seen that in the dijet case only \znunu\, + jets and W + jets events give a small 
background contribution over a clear signal.
At higher jet multiplicities $n = 3 \dots 6$, top decays as well as about one QCD event contribute 
to the remaining background after the final selection.

\subsection{Establishing a Signal incompatible with Standard Model Background in Data}
\begin{figure}
  \begin{center} 
  \subfigure[SM background only]{\label{fig:ratio_bkgd}\includegraphics[angle=90,width=7cm]{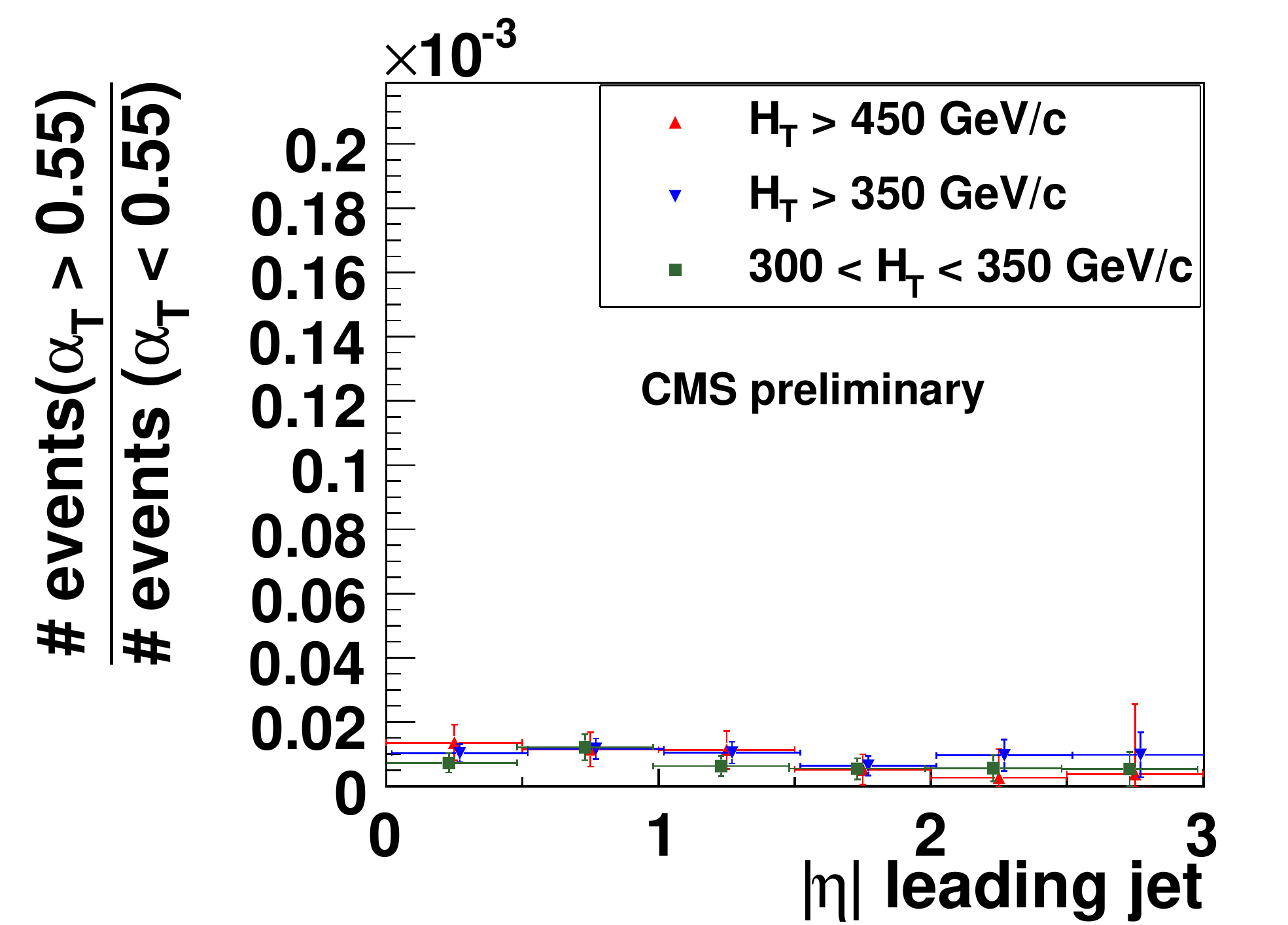}}
  \subfigure[SUSY(LM0) + SM background]{\label{fig:HT_LM0}\includegraphics[angle=90,width=7cm]{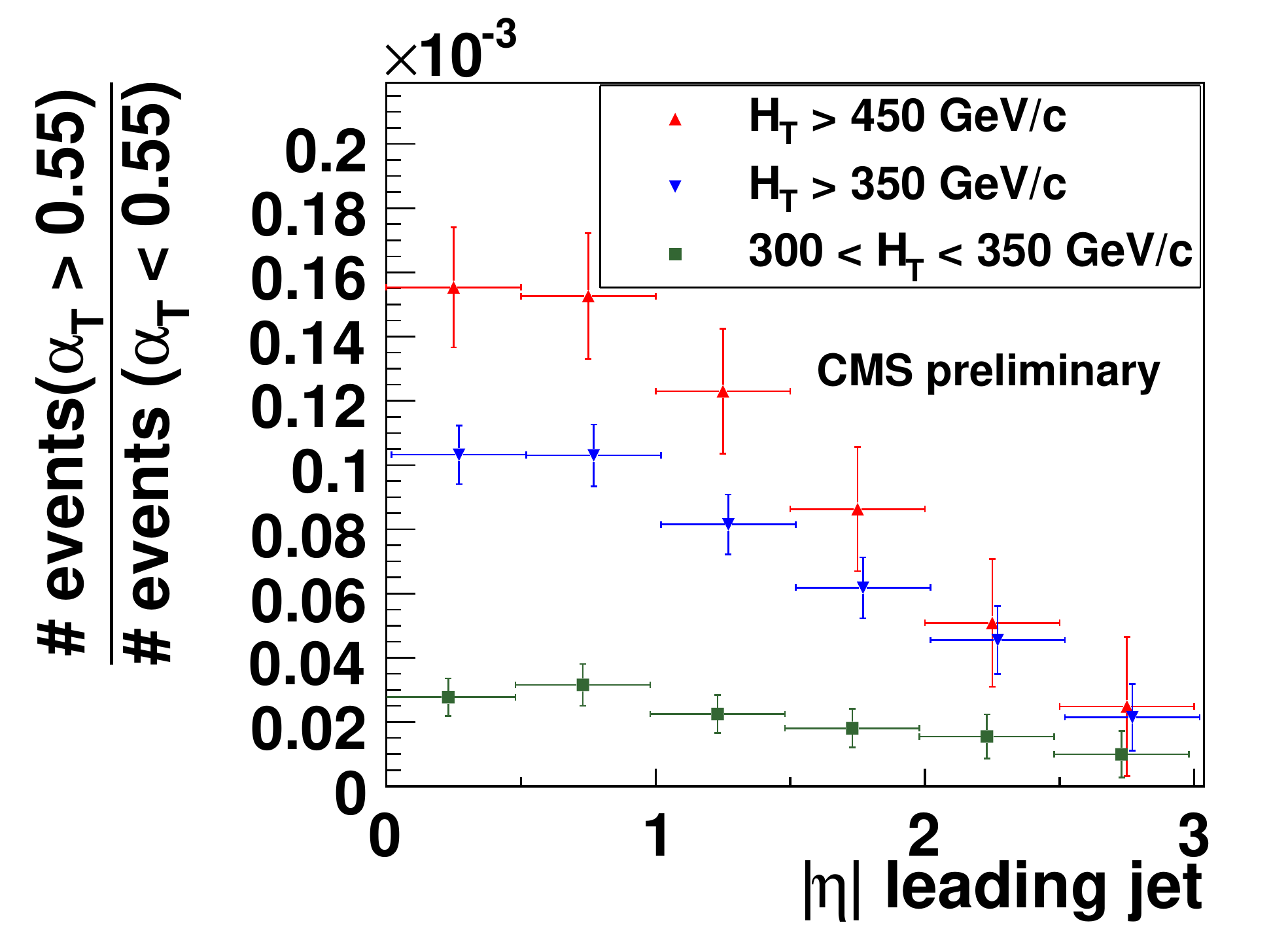}}
    \caption{
$R_{\alt}(0.55)$ as a function of $|\eta|$ for different $H_{\rm T}$ cuts for SM background only and in case of presence of a SUSY signal. The error bars indicate the expected statistical uncertainties for a data sample of 100 \mbox{\ensuremath{\,\text{pb}^\text{$-$1}}}\xspace.
    \label{fig:ratio_HTcuts2}}
  \end{center}
\end{figure}

To establish the discovery of a SUSY signal the fact that 
signal events are produced more centrally in pseudo-rapidity compared to the SM backgrounds, 
in particular compared to QCD events whose main production mechanism is $t$-channel exchange, is used. 
The pseudo-rapidity of the leading jet can be used as a measure of the centrality of an event. 
For the SM background the ratio 
$R_{\alt}(0.55)$ of events with \alt\ larger than the cut value over that of events 
with \alt\ smaller than the cut value, is, as shown in Fig.~\ref{fig:ratio_bkgd}, approximately constant as a function of 
pseudo-rapidity and independent of $H_{\rm T}$.
$R_{\alt}(0.55)$ behaves very differently in the presence of a SUSY signal as  
illustrated in Fig.~\ref{fig:HT_LM0} for the LM0 benchmark point. 
As can be seen, the presence of a SUSY signal manifests 
itself with two distinct features:
 $R_{\alpha_T}(0.55)$ exhibits a negative slope with larger values of $|\eta|$;
 tighter requirements on $H_{\rm T}$ result in a steeper slope and an offset in $R_{\alpha_T}(0.55)$.
For events with $H_{\rm T}$ $> 350$~\GeVc\ the measured
$R_{\alpha_T}(0.55)$ in the central $|\eta|$ bins is well above the
ratios obtained from the control region $300 < H_{\rm T} < 350$~\GeVc\ 
and increases with smaller values of $\eta$. Even
with a systematic uncertainty of 100\% on $R_{\alpha_T}(0.55)$ in the
control region the excess would remain convincing.

%% file: photon.tex
\section{Search for SUSY in diphoton final state}
\label{sec:phot}

The final state with two high \et\ photons and large missing transverse 
energy can harbor new physics signals in a variety of theoretical scenarios, 
most notably GMSB. 
In this section, a data-driven strategy that can be used in a 
plausible start-up scenario to predict the \MET\ distribution in a diphoton 
sample from the SM processes is described. Observation of an excess of events at high 
\MET\ would be a signature of new physics.

The SM contribution to diphoton plus \MET\ final state is small. The only 
physics backgrounds are $Z\gamma\gamma \rightarrow \nu \nu \gamma \gamma$ and 
$W\gamma\gamma \rightarrow \ell \nu \gamma \gamma$. 
The instrumental background has three major components. The first component results from QCD 
events with no true \MET\ (QCD background), such as direct diphoton, photon plus jets, and 
multijet production.  
The second component comes from events with real \MET\ (Electroweak background). 
This component is dominated by $W\gamma$ and 
$Wj$ production where the W decays into an electron plus a neutrino when 
the electron is mis-reconstructed as a photon.
The third background is associated with the high energy muons from cosmic rays 
or beam halo (Non-beam background). \\

\subsection{Trigger and selection}
The trigger applied in this analysis requires a photon with
\et\ $>$25~\GeV. This trigger is expected to run unprescaled for the 
duration of the first CMS run, and is fully efficient for the SUSY signal considered. 

For the photon candidate selection, two objects in the ECAL Barrel ($|\eta| \leq 1.45$)
with \pt\ $>$30~\GeVc\ were required to be isolated and to have a negligible deposit of energy
in the hadronic calorimeter around the photon.
The selected objects are classified as electrons if they have an associated 
track stub in the pixel detector, referred to as pixel seed, and as photons otherwise.
In this way three independent samples are defined: {\it$\gamma \gamma$} sample comprising events with at east two
photons; {\it $e\gamma$} sample with at 
least one electron and  at least one photon; {\it $ee$ } sample with at least two electrons.

Finally a cut \MET\ $>$ 80~\GeV\ is applied to reduce significantly the SM background. 
The results of the selection are reported in Tab.~\ref{tab:count1}

\begin{table}[h]
\begin{center}
\begin{tabular}{|l||c|c|c|c|c||c|}
\hline
 &$\gamma\gamma$ &$\gamma$\small{+jet}  &\small{multijets}   &$W$, $W$\small{+n}$\gamma$ &$Z$, $Z$\small{+n}$\gamma$ & \small{GM1c}\\
\hline \hline
\small{all events} & 1055 $\pm$ 4 & 3189 $\pm$ 100 & 173 $\pm$ 37  &  8.5 $\pm$ 3.0  &23$\pm$ 1.3   &20.7 \\ \hline
\small{after 
\MET}
 & 1.3 $\pm$ 0.16  & 1.3 $\pm$ 0.16 & &0.09$\pm$ 0.04 & 0.09$\pm$ 0.02 &14.8 \\ \hline
\end{tabular}
\caption{\label{tab:count1} Event counts before and after the cut on \MET\ for the $\gamma\gamma$ sample for  100\mbox{\ensuremath{\,\text{pb}^\text{$-$1}}}\xspace at 10 \TeV. }
\end{center}
\end{table}

\subsection{Non-beam background}
Energetic cosmic muons or muons from beam halo can emit photons as they pass 
through the ECAL. The shower shape and arrival time of these photons is slightly 
different from the ones originating from the interaction point, especially
for beam halo.

\begin{figure}
\begin{center}
\subfigure[Photon shower width along $\eta$ versus 
photon $\eta$]{\label{fig_halo_etaWidth_2D}\includegraphics[width=7cm]{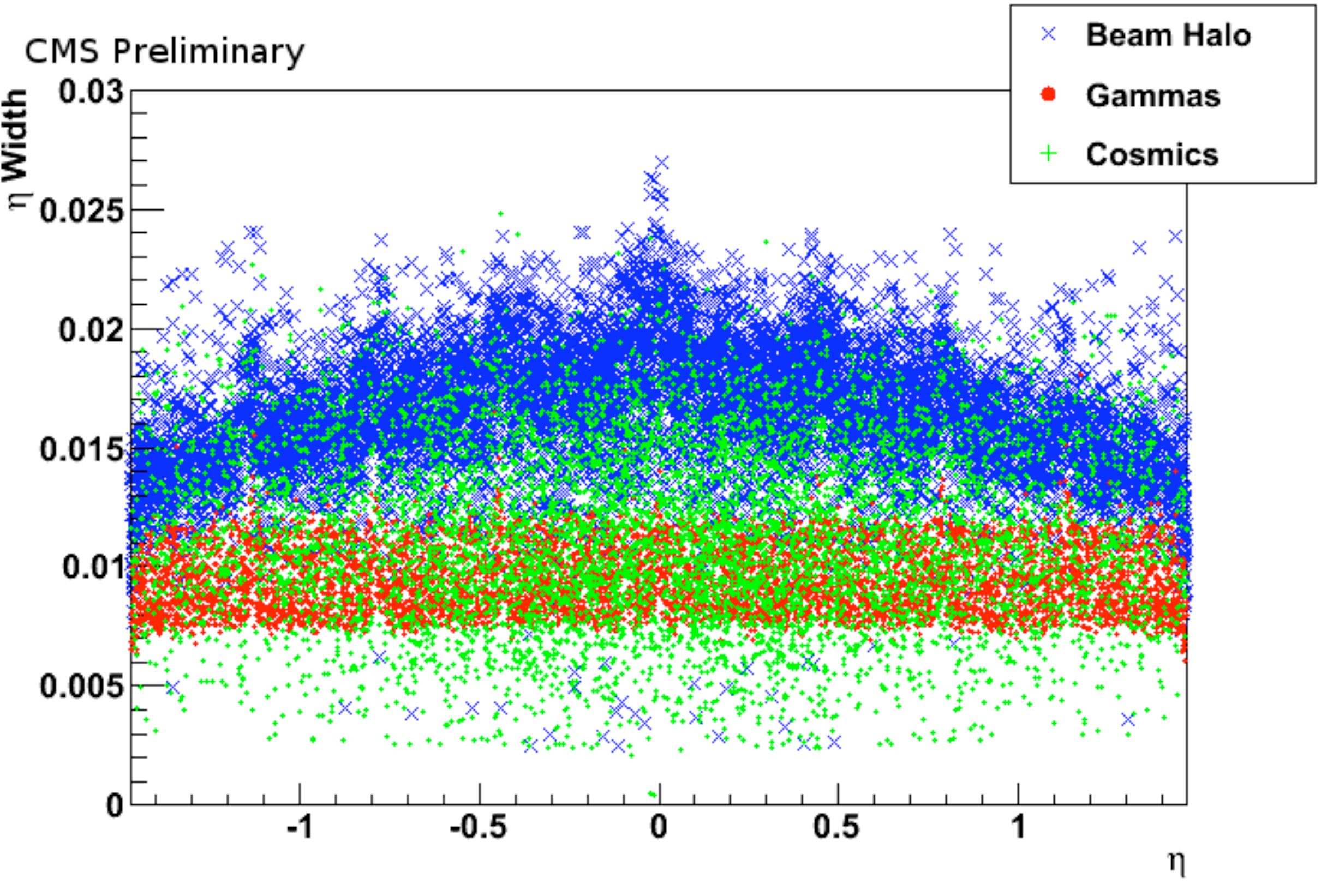}}
\subfigure[Measured cluster time versus $\eta$]
{\label{fig_halo_2d}\includegraphics[width=7cm]{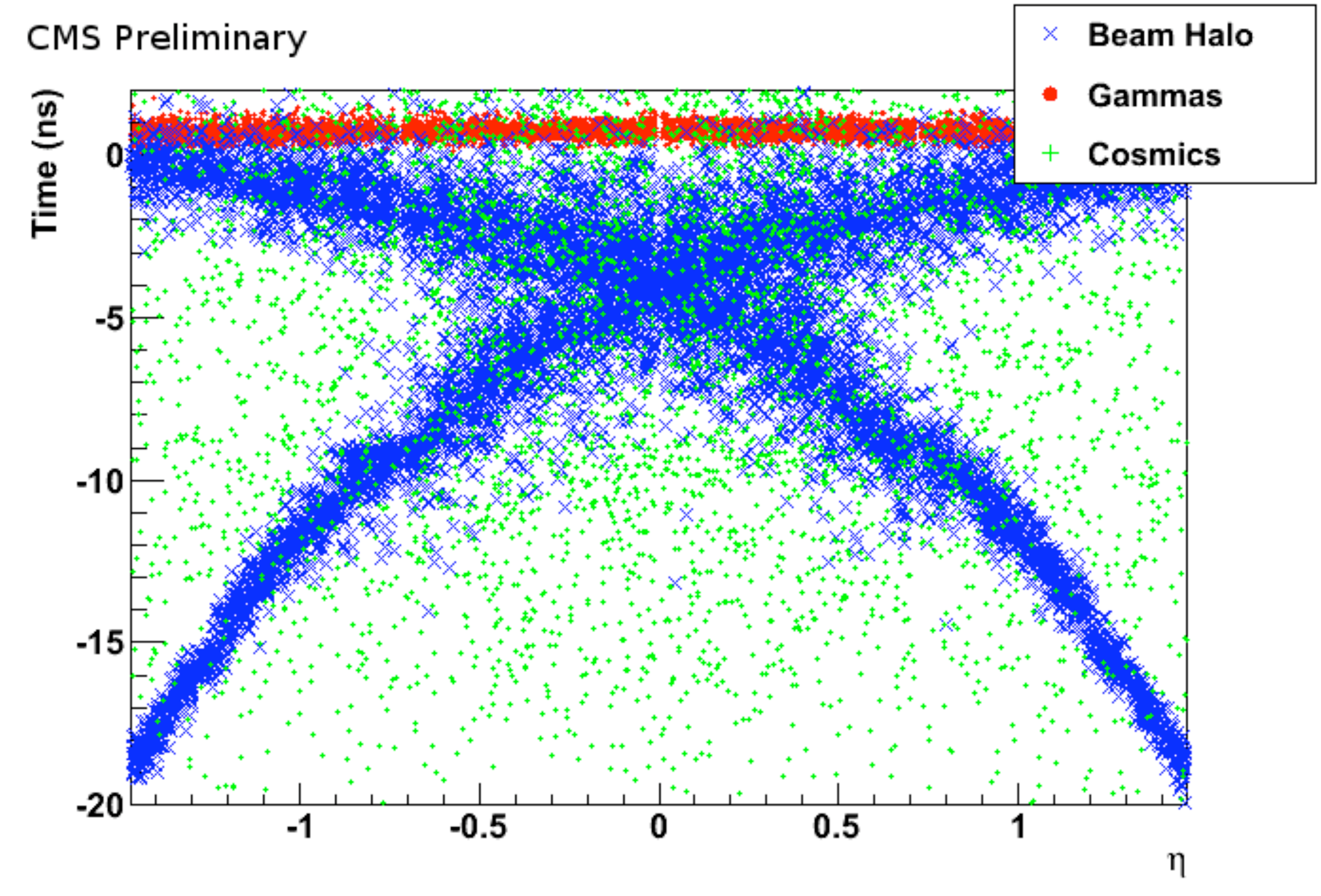}}
\caption{\label{fig:nonbeam} Photon shower and Timing in beam halo, cosmics, and photon gun events.}
\end{center}
\end{figure}

Photons from the interaction point hit a crystal approximately 
perpendicular to the face because the crystals are rotated in $\eta$ such 
that they point back to the interaction point. 
As a result of this rotation, photons from the interaction point give rise
to showers with a small spread in $\eta$. 
However, photons from the beam halo traveling parallel 
to the beam tend to have a large spread in $\eta$. Cosmic ray photons
come from all possible angles and could have very narrow or very wide showers.
Fig. \ref{fig_halo_etaWidth_2D} displays the distribution of the energy-weighted
RMS of the shower profile in $\eta$ ($\eta$-width) for the three sources of photons.

The expected time resolution for individual crystals is better then 1 ns,
and the non-beam backgrounds can be discriminated against using the
measured cluster time. Cosmic background is asynchronous, 
but beam halo has a very specific time distribution. Protons and 
halo muons travel parallel to each other and have the same $z(t)$. The time 
difference between the photons from the IP and halo is then given by
$\Delta t  =  (Z+\sqrt{Z^2+R^2}) / c$.
The value of R, the radius of the shower in the ECAL, is not known. 
Fortunately, the resulting uncertainty in 
$\Delta t$ is fairly small, as shown in Fig. \ref{fig_halo_2d}, which shows 
the comparison of the measured time of the photons from beam halo, cosmics, and prompt photons,
corrected for the detector geometry. 

Although the exact amount of non-beam backgrounds is very hard to predict, 
a combination of the two techniques explored above (shower shape and timing) 
will be sufficient for the effective elimination of these backgrounds.
It will also be possible to isolate a pure sample of non-beam backgrounds, 
ensuring a reliable estimation of the residual contamination.

\subsection{Determination of \MET\ Distribution from Data} 
The contribution of the Electroweak background is determined using the 
$e\gamma$ sample. 
The \MET\ distribution in {\bf $e\gamma$} sample is multiplied by 
$f_{e\rightarrow\gamma}/(1-f_{e\rightarrow\gamma})$, 
where $f_{e\rightarrow\gamma}$ is the electron-photon mis-identification rate.
$f_{e\rightarrow\gamma}$ is obtained by fitting the 
mass of the Z for the $ee$ sample and $e\gamma$ sample, and by comparing the 
integrals of these fits.
The resultant distribution is taken as an estimate of the 
this background and is subtracted prior to assessing the background from QCD events.\\
The \MET\ distribution for the  QCD background is modelled using a 
sample of Z$\rightarrow ee$ events, selected from $ee$ sample with
invariant mass cut on the two electrons, 80 $< M_{ee} <$ 100~\GeV.

The key assumption is that
the di-EM system which is measured comparitively well, and the recoil and other
hadronic activity which is measured poorly can be separated in the event.
The di-EM $p_T$ is used as a measure of this additional activity.
Its amount is different for the Z$\rightarrow ee$ and the {$\gamma\gamma$}
events. To obtain the proper shape of the \MET\ distribution,
the Z$\rightarrow ee$ events are reweighted so that their di-EM $p_T$
distribution matches the one in the {$\gamma\gamma$} sample.
The weighted \MET\ distribution is then normalized to the 
{$\gamma\gamma$} sample in the
low missing transverse energy (\MET\ $<$ 20~\GeV) region.

\begin{figure}
\begin{center}
\subfigure[SM background only]{
	\label{close-sm-zee}\includegraphics[width=7cm]{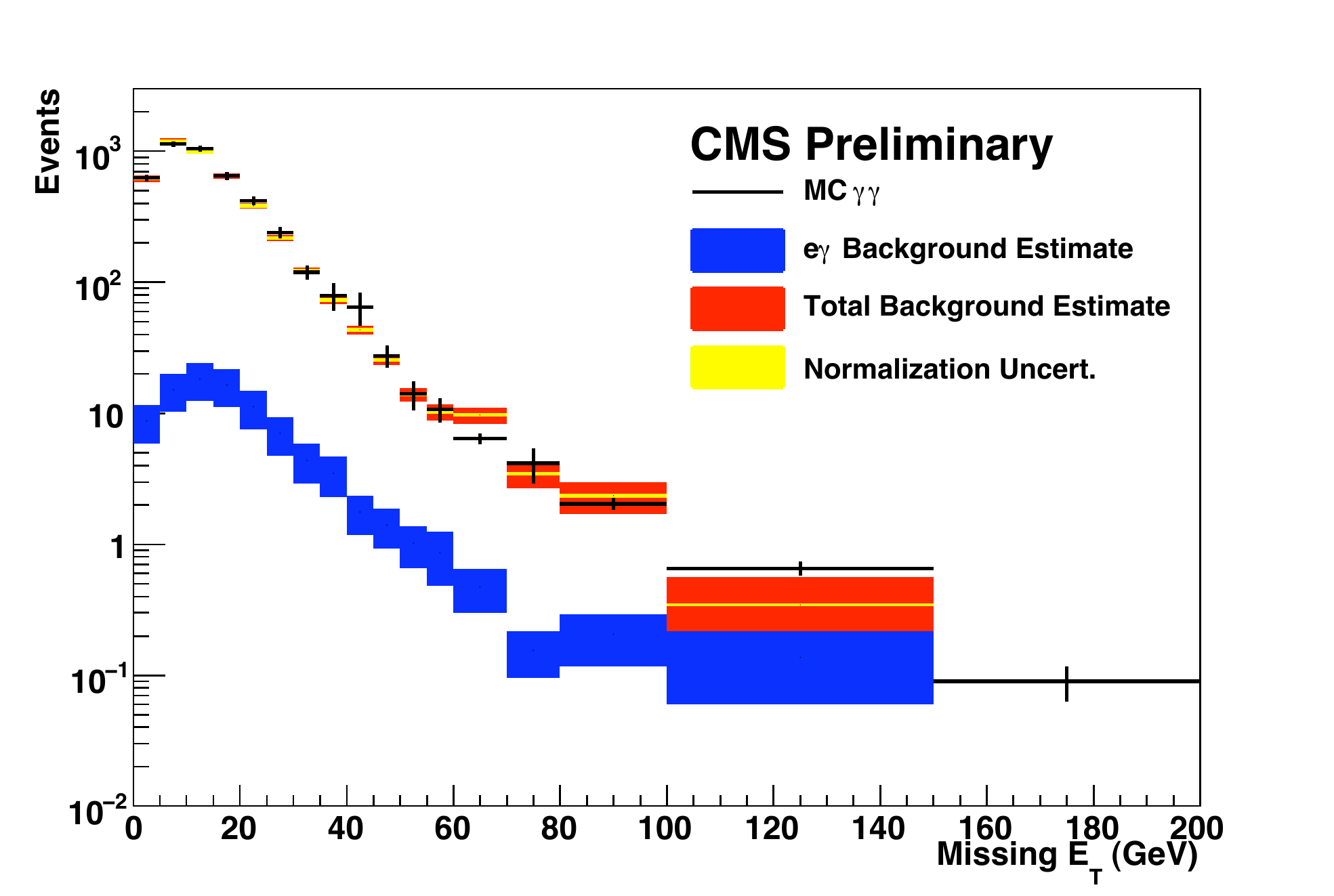}}
\subfigure[SUSY(GM1c) + SM background]{
	\label{close-gm1c-zee}\includegraphics[width=7cm]{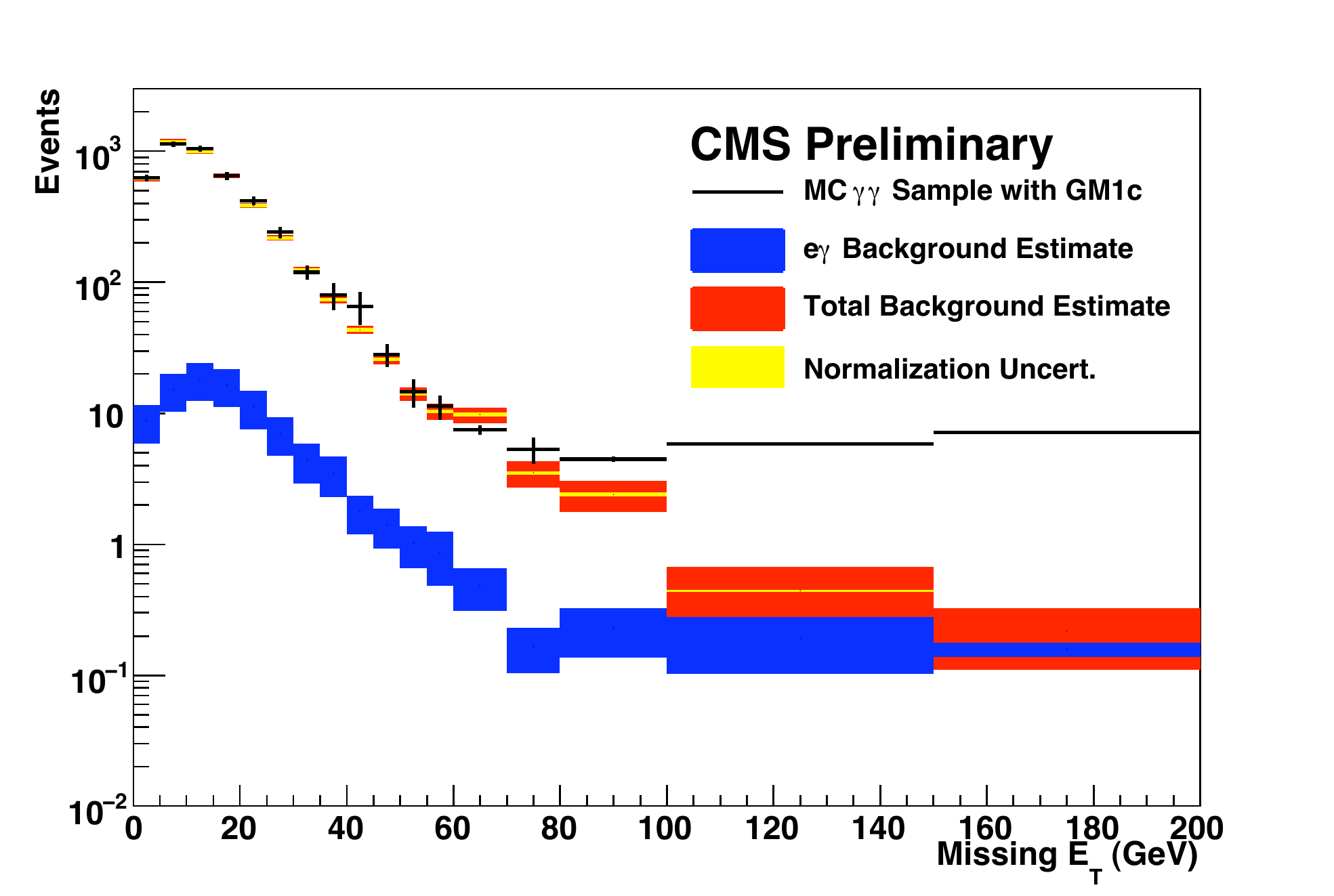}}	
 \caption{Background closure test using 
$Z\rightarrow e^+e^-$ events to describe the QCD background.}
\end{center}
\end{figure}

Fig. \ref{close-sm-zee} show the closure test for 
$Z\rightarrow e^+e^-$ sample used for QCD background determination in absence of SUSY signal.
The presence of the signal can bias the background estimation, thus the entire analysis is repeated
with GM1c SUSY signal mixed in, as shown in Fig. \ref{close-gm1c-zee}.

The data driven estimation of the background agrees very well with the number of expected events 
evaluated with the MC truth, as shown in Tab.~\ref{tab:close}.
\begin{table}
\begin{center}
\begin{tabular}{|l||c|c|c||c|c|}
	\hline
 & \multicolumn{3}{|c||}{MC truth}&  \multicolumn{2}{|c|}{Data-driven} \\ \hline
 &$N^{QCD}_{\gamma\gamma}$ &$N^{EWK}_{\gamma\gamma}$ & $N^{\sc GM1c}$&$N^{QCD}_{BG}$  &$N^{EWK}_{BG}$   \\ \hline \hline
no SUSY & 2.61 $\pm$ 0.23 &0.17 $\pm$ 0.04 &   & 2.34 $\pm$ 0.65&  0.35 $\pm$ 0.10 \\ \hline
with SUSY &  2.61 $\pm$ 0.23&  0.17 $\pm$ 0.04 &14.8 $\pm$ 0.1 & 2.48 $\pm$ 0.67  & 0.50 $\pm$ 0.10 \\ \hline
\end{tabular}
\caption{\label{tab:close}  Closure test for $\gamma\gamma$ sample with and without {\sc GM1c} SUSY signal. 
 The number of events corresponds to 100 \pbinv\ at 10 \TeV.} 
\end{center}
\end{table}

%% file: lepton.tex
\section{SUSY discovery potential and measurement of a dilepton mass edge}
\label{sec:lept}
In mSUGRA models the lightest neutralino escapes 
detection and no mass peaks can be observed in SUSY decay chains. Of 
special interest are robust signatures such as edges in mass distributions 
in leptonic final states which can be probed with the CMS experiment.

The purpose of this analysis is to observe a significant excess of opposite 
sign same flavour leptons over the various backgrounds and to measure the endpoint 
in the invariant mass distribution. All flavour symmetric background (including SUSY 
decays of this type) can be determined from data events with opposite sign 
opposite flavour leptons. The aim is to perform such an analysis already with 
the first LHC data which is expected to amount to roughly 200-300\pbinv\ in 2010. 

The leptonic decay of the next-to-lightest neutralino gives a characteristic 
signature. This decay can proceed in different ways even in the mSUGRA model. 
A mass difference of the neutralinos smaller than the Z boson mass and any 
slepton mass leads to a three body decay. In that case the endpoint in the lepton 
invariant mass represents directly the mass difference of the two lightest neutralinos
\begin{equation}
m_{ll,max} = m_{\tilde{\chi}_2^0} -m_{\tilde{\chi}_1^0}
\label{eq:neutralino3body}
\end{equation}
A two-body decay occurs via a real slepton and is allowed if at least one 
slepton is lighter than the mass difference of the neutralinos. In that 
case the endpoint can be expressed by
\begin{equation}
\left( m_{ll}^{max} \right)^2 = \frac{\left( m_{\tilde{\chi}_2^0}^2 - m_{\tilde{l}}^2 
\right) \left( m_{\tilde{l}}^2 -m_{\tilde{\chi}_1^0}^2\right)}{m_{\tilde{l}}^2}
\label{eq:neutralino2body}
\end{equation}
where $m_{\tilde{l}}$ is the mass of the intermediate slepton. The shape of the mass 
edge results only from kinematics and is triangular.

\subsection{Trigger and selection}
Single lepton (electron or muon) trigger is used in this analysis.
Since the leptons originating from the signal decay have a very soft 
\pt\ spectrum, the triggers with the lowest 
available threshold for electrons (\pt\ $>$ 15~\GeVc) and muons (\pt\ $>$ 11~\GeVc) are used.

The base selection requires two isolated leptons of opposite sign.
Muon identification requires reconstruction in both the muon system and the inner 
tracker~\cite{muonReco}. Each electron has to fulfill the tight electron 
identification criteria, which consist of a set of cuts depending on the 
electron \pt\ and $\eta$~\cite{PASelectron}. 
Additionally  a \pt\ $>$ 10~\GeVc\ and  
$|\eta|<2$ is required for each lepton. 
To stay above the trigger threshold the first lepton is required to have 
a \pt\ $>$ 16~\GeVc. The main SUSY selection is based on jets and 
missing transverse energy. The cuts have not been 
optimized at a certain benchmark point, but should reflect 
the general SUSY signature. The selection requires three  
jets with \ptjf~$>$~100~\GeVc, \ptjs~$>$~50~\GeVc, 
and \ptjt~$>$~50~\GeVc. A  missing 
transverse energy of at least 100~\GeV\ is required.

\begin{table}[hbtp]
\begin{center}
\begin{tabular}{|l||c|c|c|c|c||c|c|}
\hline
Sample  &tt+jets &Z+jets &W+jets &Diboson &Dijet&LM0 signal &LM0 incl.  \\
\hline \hline
N$_{events}$&80 &1 &2 &0 &0 &87 & 362 \\
\hline
\end{tabular}
\caption{\label{tab:Cutflow}Number of selected events using the described event selection 
for an integrated luminosity of 200 \mbox{\ensuremath{\,\text{pb}^\text{$-$1}}}\xspace.}
\end{center}
\end{table}
The number of events obtained after the selection is listed in Tab.~\ref{tab:Cutflow}. 
All background which leads to uncorrelated lepton pairs 
can be measured directly from data~\cite{georgia}. 
In order to extrapolate to the same flavour opposite sign lepton pair distribution
the opposite sign opposite flavour lepton pairs and use this distribution are selected. 
With this method one is able to predict all backgrounds which produce uncorrelated 
leptons such as $W$, $t\bar{t}$, dijet and $WW$ events.

\begin{figure}[t]
   \begin{center}
\subfigure[$ee$, $\mu\mu$ ]{\label{fig:invmsig}\includegraphics[angle=90,width=7cm]{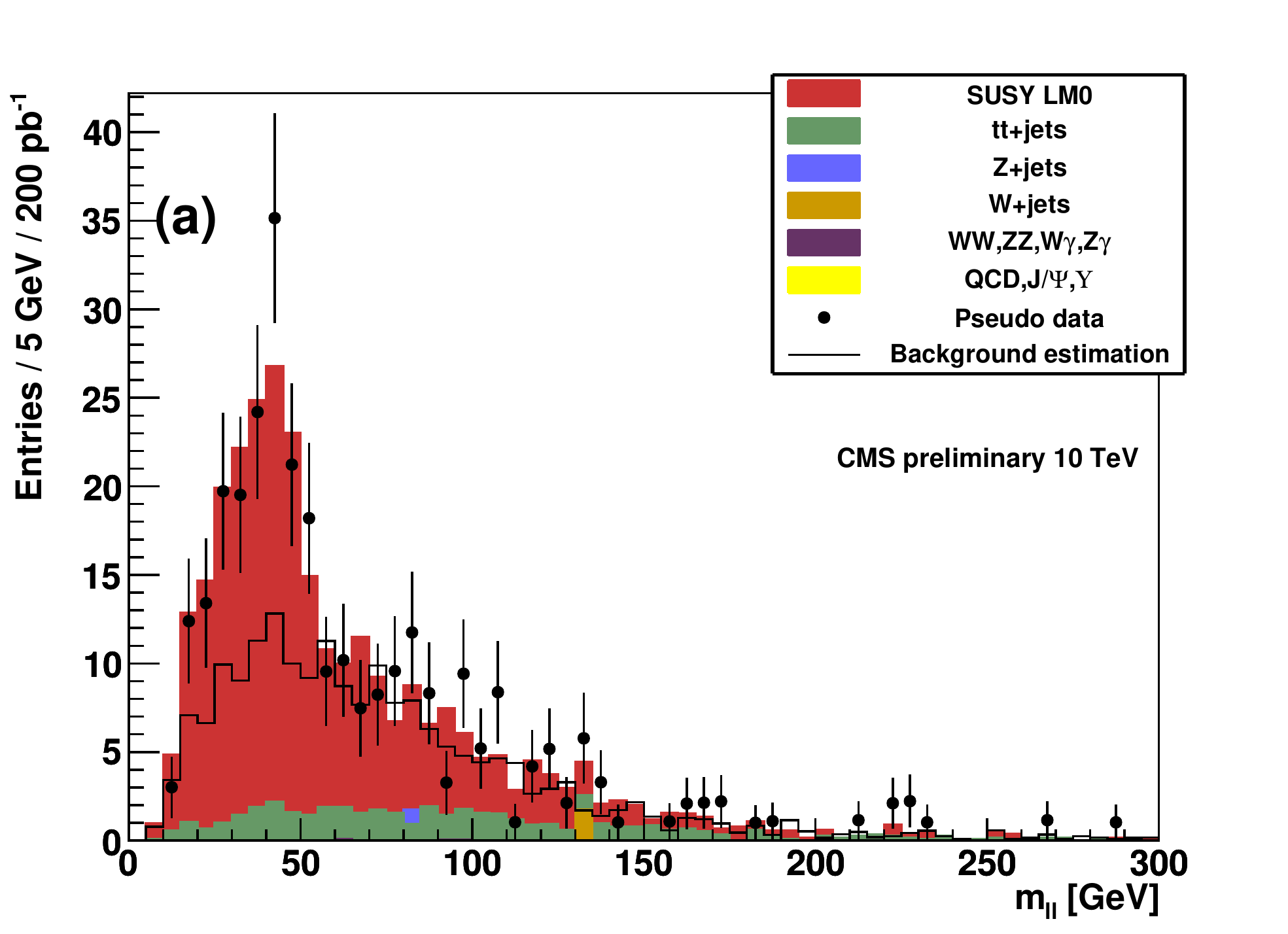}}
\subfigure[$e\mu$]{\label{fig:invmbkg}\includegraphics[angle=90,width=7cm]{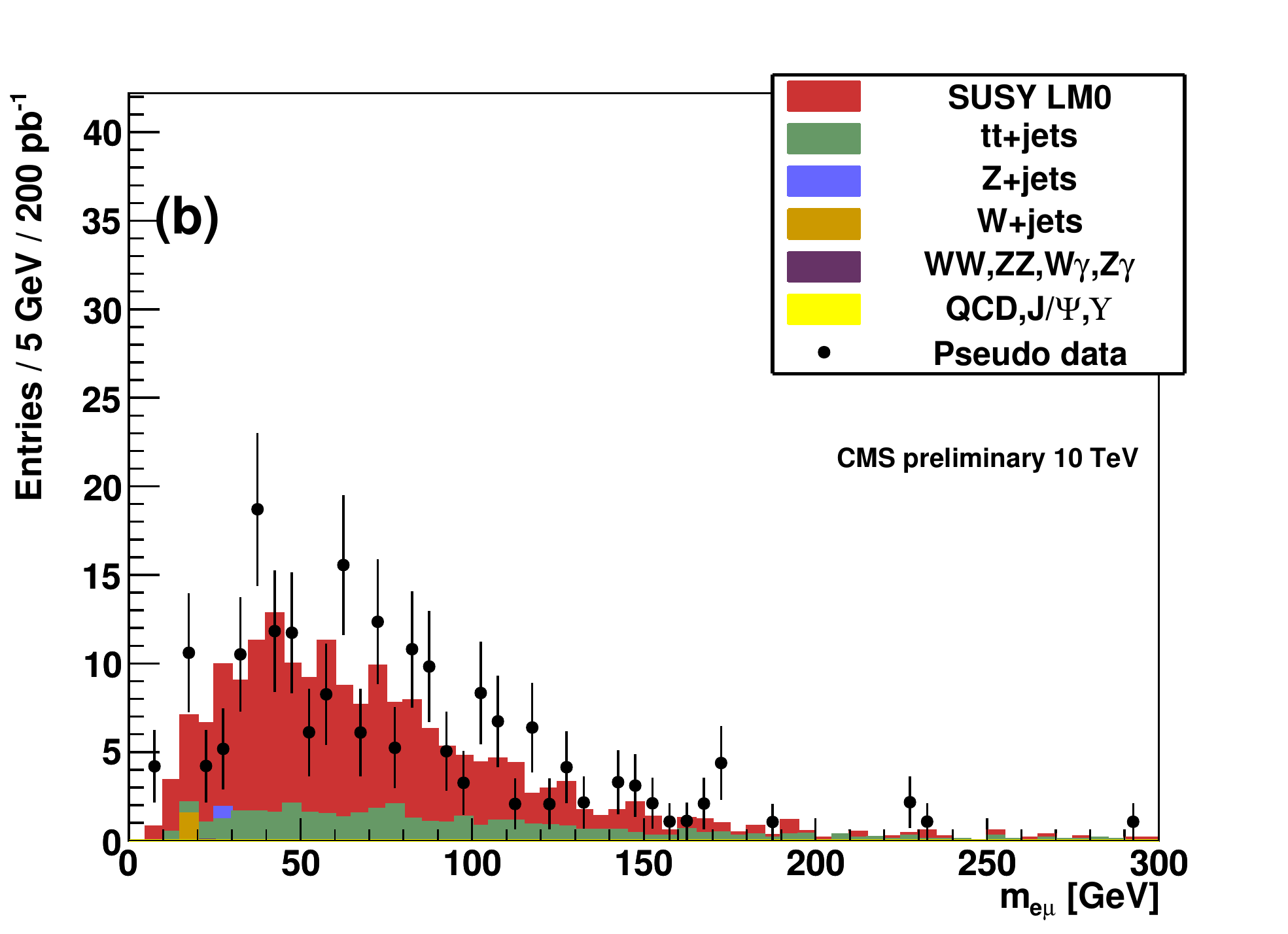}}
\end{center}
 \caption{\label{fig:SFOS_with_cuts}DiLepton invariant mass. 
The black solid line represents the extrapolation from the opposite 
flavour distribution. The black points represent pseudo data of one experiment
where no scaling has been applied but exactly 200 \pbinv\ of 
MC events have been analysed.}
\end{figure}
The invariant mass distribution of all opposite sign same 
flavour leptons for 200\pbinv\ is shown in Fig.~\ref{fig:invmsig}. 
The opposite sign opposite flavour distribution used to extrapolate the background 
is displayed in Fig.~\ref{fig:invmbkg}.

\subsection{Determination of the mass edge}
\label{sec:fit}

The model used for the fit of the mass edge consists of three parts. To model the signal the theoretical model~\cite{shapedependence,theomodel} convoluted numerically with a gaussian has been used in case of a 3-body decay
\begin{eqnarray}\label{eq:fit_func}
S(m_{ll}) = \frac{1}{\sqrt{2\pi}\sigma} \int\limits_{0}^{m_{cut}} dy \cdot 
 y \frac{\sqrt{y^4-y^2\left(m^2+M^2\right)+\left(m\,M\right)^2}}{\left(y^2-m_Z^2\right)^2} \nonumber \\
       \times 
\left( -2 y^4-y^2\left(m^2+2M^2\right)+\left(m\,M\right)^2 \right)
e^{\frac{-\left( m_{ll}- y \right)^2}{2\sigma^2}}
\end{eqnarray}
where $m=m_{\tilde{\chi}_2^0}-m_{\tilde{\chi}_1^0}$ is the difference, $M = m_{\tilde{\chi}_2^0}+m_{\tilde{\chi}_1^0}$ is the sum in neutralino mass and $M_Z$ is the Z mass, which is kept fixed.
In case of the two-body decay the signal model consists of a triangle convoluted with a gaussian  
\begin{equation}\label{eq:fit_func2}
T(m_{ll}) = \frac{1}{\sqrt{2\pi}\sigma} \int\limits_{0}^{m_{cut}} dy \cdot y e^{\frac{-\left( m_{ll}- y \right)^2}{2\sigma^2}}
\end{equation}
A curve parametrized as
\begin{equation}\label{eq:fit_bkg}
B(m_{ll}) = m_{ll}^{a} \cdot e^{-b\cdot m_{ll}}
\end{equation}
has been used to fit the opposite sign opposite flavour invariant mass distribution. 
Additionally the Z peak is fitted using a Breit-Wigner convoluted with a gaussian.
The number of signal $N_{Sig}$, background $N_{Bkg}$ and Z events $N_Z$ are fitted as well. 

\subsection{Expected results}
A simultaneous fit to the the $ee$, $\mu\mu$ (signal plus background model) 
and $e\mu$ (background model)  invariant mass distributions is performed.
\begin{figure}[t]
   \begin{center}
\subfigure[$ee$, $\mu\mu$ invariant mass distributions]{\label{fig:fitsig}\includegraphics[angle=90,width=7cm]{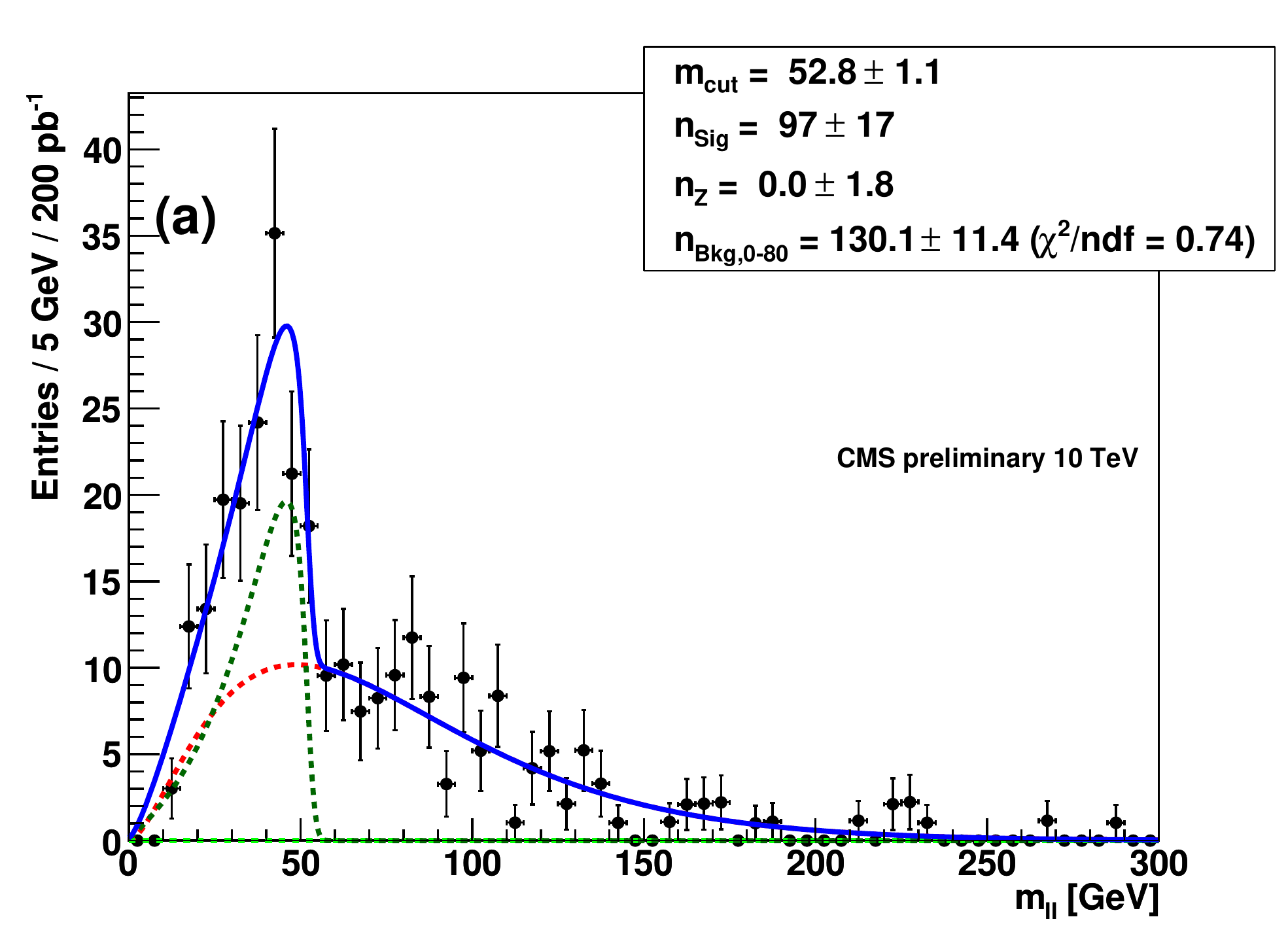}}
\subfigure[$e\mu$ invariant mass distribution]{\label{fig:fitbkg}\includegraphics[angle=90,width=7cm]{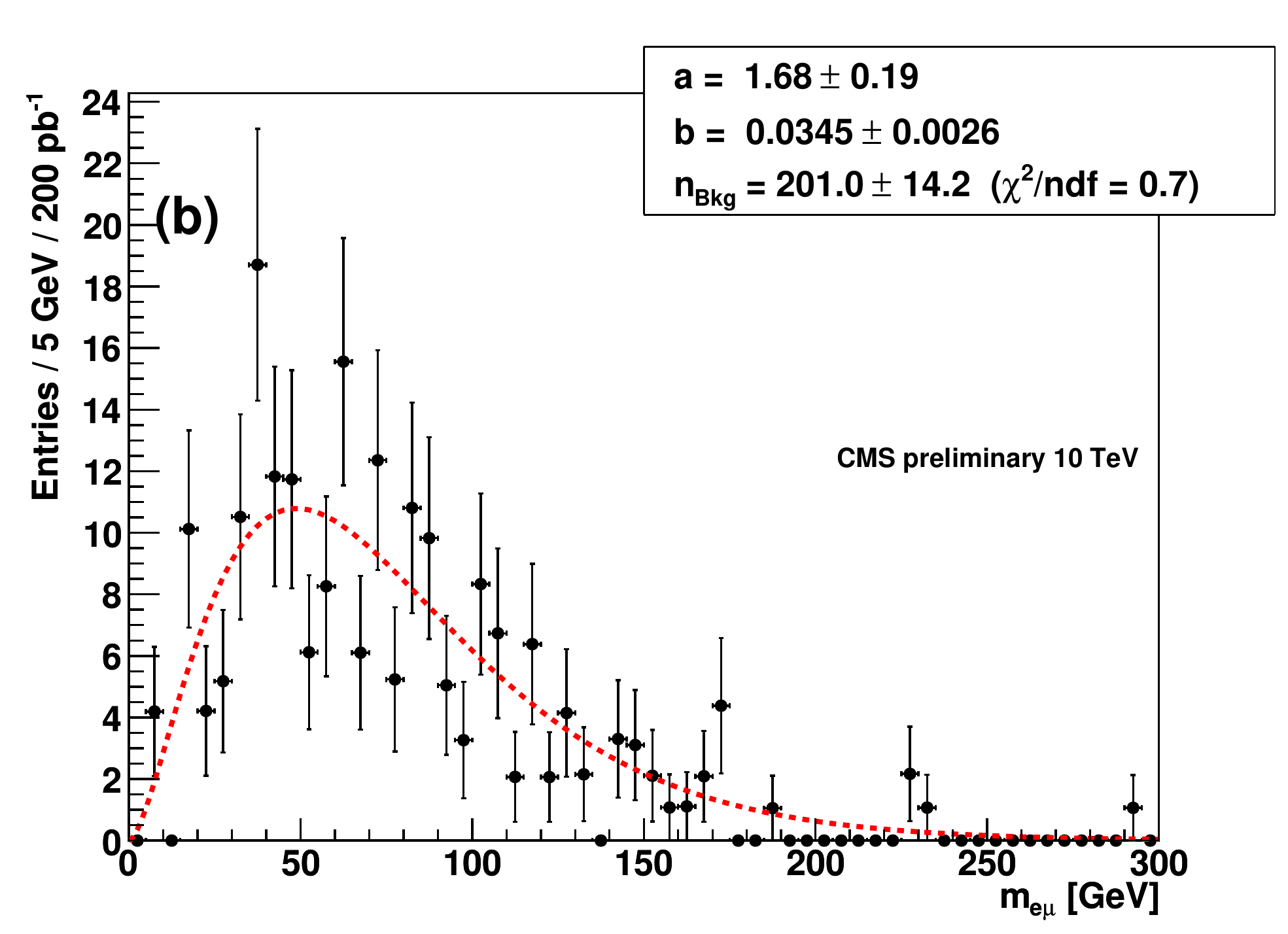}}
\end{center}
 \caption{\label{fig:fit_LM0} The combined fit at LM0, 
the green curve represents the SUSY signal model, 
the red curve is the background function and the 
light green dashed line the Z contribution. }
\end{figure}

The fit to the invariant mass distributions at LM0  is 
shown in Fig.~\ref{fig:fitsig} and it yields
 a value of $ m_{ll,max} = \left( 52.8 \pm 1.1 \right) \textnormal{$ \GeV$}$
for the dataset of exactly 200 \pbinv\
and  $ m_{ll,max} = \left( 51.3 \pm 1.5 \right) \textnormal{$ \GeV$}$ for the 
full MC data where the error is the one expected after 200 \pbinv. 
The derived number of signal events $n_{sig} = 97 \pm 17$ agrees with the number of signal events from 
MC truth (Tab.~\ref{tab:Cutflow}). The theoretical endpoint $m_{ll,theo}=52.7$~\GeV\ 
is reproduced in case of the fit with the three-body decay model.
The background fit of the opposite sign opposite flavour lepton pairs is shown in Fig.~\ref{fig:fitbkg}.
The total number of background events is $n_{Bkg} = 201 \pm 14$
which is in agreement with the expected number from MC truth (192).

The main sources of systematic uncertainties are the jet energy scale, the electron energy scale, 
the lepton efficiency and the modeling of the background and of the resolution. 
Combining systematic and statistical errors, the expected results for LM0 and 200\pbinv\ at 10 \TeV\ is
\begin{equation}
    m_{ll,max} = \left( 51.3 \pm 1.5_{stat.} \pm 0.9_{syst.} \right) \textnormal
{\ensuremath{{\,\text{Ge\hspace{-.08em}V\hspace{-0.16em}/\hspace{-0.08em}c} }} }
\end{equation}
compared to a theoretical value of $52.7$~\GeV.

%% file: conc.tex
\section{Summary}
The strategy for three different analyses aimed to discovery SUSY 
in the first year of data-taking at LHC considering $\sqrt{s}=~10\TeV$ with the CMS experiment were presented. \\

The search for SUSY in multijet final state is carried out in the context of 
SUSY for several sets of parameters in the mSUGRA
parameter space assuming an integrated luminosity of 100~\pbinv . 
The discrimination power of \alt\ against SM background from QCD
events provides, for favourable SUSY benchmark points, signal over background ratios of 4 to
8 depending on the considered jet multiplicity bin. 

In the context of the GMSB models a strategy to infer the existence of new physics 
in diphoton events with large missing transverse energy has been developed.
Background contribution in the high \MET\ region is estimated can be precisely estimated
with  a pure data-driven method in  100~\pbinv .

A significant excess of SUSY opposite sign same flavour lepton pairs can be found within the
first 200~\pbinv\ at LM0. The signal provides a quite robust signature and the background determination
directly from data is possible.
At LM0 the combined fit of the dileptonic endpoint is possible with
200 \pbinv with an expected uncertainty of 1.8\GeV.